\documentstyle[twoside, epsfig]{article}

\input ibvs2.sty

\begin{document}
\IBVShead{6145}{00 Month 200x}

%\IBVStitlel{The visual binary HIP10680/HIP10679 in the $\beta$ Pictoris Association}{with one of the fastest members.}

\IBVStitletl{HIP10680/HIP10679: a visual binary in the $\beta$ Pictoris Association}{ with the fastest rotating  member.}

\IBVSauth{Messina, S.$^1$; Hentunen V.-P.$^2$, Zambelli, R.$^3$}

\IBVSinst{INAF- Catania Astrophysical Observatory, via S.Sofia, 78 I-95123 Catania, Italy, e.mail: sergio.messina@oact.inaf.it}
\IBVSinst{Taurus Hill Observatory,  Varkaus, Finland, e-mail: veli-pekka.hentunen@kassiopeia.net}
\IBVSinst{Canis Mayor Observatory, La Spezia, Italy, e-mail: robertozambelli.rz@libero.it }

\SIMBADobjAlias{HD14082}{BD+28 382}{SAO75265}{TYC 1777-1479-1}
\IBVStyp{F5V+G2V}
\IBVSkey{photometry}
\IBVSabs{We present the results of a multi-filter photometric monitoring of the wide binary HIP10680/HIP10679}
\IBVSabs{WE found both component to be variable with amplitude up to $\Delta$ V = 0.03 mag
in the case of HIP10680 and $\Delta$ V = 0.07 mag in the case of HIP10679.}
\IBVSabs{We could measure the rotation periods P = 0.2396\,d of the hotter F5V component HIP10680 and P = 0.777\,d of the cooler 
G5V component HIP10679}
\IBVSabs{We found that the rotation axes of both components are aligned with an inclination $i$ = $\sim$ 10$^{\circ}$.}

\begintext

\section*{Introduction}
We are carrying out a photometric monitoring of confirmed and candidate members of the young $\beta$ Pictoris
Association. Particular emphasis is given to multiple stellar systems to study the distribution of the rotation periods
of their components. We want to investigate what causes significant differences among the rotation periods.
Causes can be either different initial rotation periods or primordial disc lifetimes. Specifically, we find that components
with very close either stellar or sub-stellar mass companions tend to exhibit a rotation period shorter than more distant
components  (see, e.g. Messina et al. 2014, 2015). \rm
In this paper, we present the case of the wide visual binary   HIP\,10680/HIP\,10679 \rm for which we have measured for
the first time the rotation periods.

\section*{Literature information}
HIP\,10680 (RA = 02:17:25.3, DEC = +28:44:42.1, J2000, V = 6.95\,mag) and HIP\,10679 (RA = 02:17:24.73, DEC =  +28:44:30.3, J2000, V = 7.75\,mag)
are components of a common proper motion visual binary (also named HD\,14082AB, BD+28 382AB) consisting of two  F5V + G2V dwarfs.
An angular separation $\rho$ = 13.8$^{\prime\prime}$ between the two components is reported in The Washington Visual 
Double Star Catalog (Mason et al. 2001). The parallaxes measured by Hipparcos have an uncertainty of the order of 15\%, and correspond to
distances  d = 34.5\,pc for HIP\,10680 
and  d = 27.3\,pc  for HIP\,10679. The most reliable distance determination was recently provided by Pecaut \& Mamajek (2013),
who report for both components a kinematic distance d = 37.62$\pm$2.73\,pc. This measurement is based on UCAC4 
proper motions (Zacharias et al. 2013), the assumption of membership to the $\beta$ Pictoris association, and the use 
the convergent point solution. In fact, this visual binary system is a well known member of $\beta$ Pictoris. 
Its membership was first proposed by Zuckerman \& Song (2004), and subsequently confirmed by  Torres et al.
(2006),   L\'epine \& Simon (2009), \rm Kiss et al. (2011), and more recently by  Malo et al. (2014).

The cooler G2V component HIP\,10679 hosts a debris disc first detected   based on its infrared excess using the MIPS (Multiband Imaging Photometer for
Spitzer) instrument onboard the 
Spitzer Space Telescope (Rebull et al., 2008). \rm They derived a disc
radius of 20\,AU and a luminosity ratio L$_{\rm d}$/L$_{\star}$ =  80$\times$10$^{-5}$. The disc
was subsequently detected  by   Herschel Space Observatory, 
whose observations allowed Riviere-Marichalar et al. (2014) \rm  to  infer an inner radius of 8.5\,AU, 
mass 3.7$\times$10$^{-3}$M$_{\oplus}$, and T$_{\rm dust}$ = 97\, K. In contrast, the same observations did not detect
any evidence for disc around the hotter F5V component HIP10680. Both components were observed by Brandt et al. (2014) as part of the SEEDS 
 high-contrast imaging survey of exoplanets and disks, but no companion was detected  within
a projected separation of 7.5$^{\prime\prime}$ ($\sim$210\,AU).

HIP\,10680 and HIP\,10679 have projected equatorial velocities  $v \sin{i}$ = 37.6 kms$^{-1}$  and $v \sin{i}$ = 7.8 kms$^{-1}$, respectively (Valenti \& Fisher 2005). 
Similar values,  $v \sin{i}$ = 45 kms$^{-1}$  and $v \sin{i}$ = 8 kms$^{-1}$, respectively, are measured by Torres et al. (2006).
Both components have well detected Li line.  Mentuch et al. (2008) measured  EW = 132\,m\AA\,\, and EW = 172\,m\AA\,\, for HIP\,10680 and HIP\,10679, respectively; da Silva et al. (2009) measured EW = 140\,m\AA\,\, and EW = 160\,m\AA\,\, for HIP\,10680 and HIP\,10679, respectively. Fast rotation and high lithium content are indicators of youth and
are well consistent with the young age of 23 Myr inferred by Mamajek \& Bell (2014) for the $\beta$ Pictoris association. 

\IBVSfig{15cm}{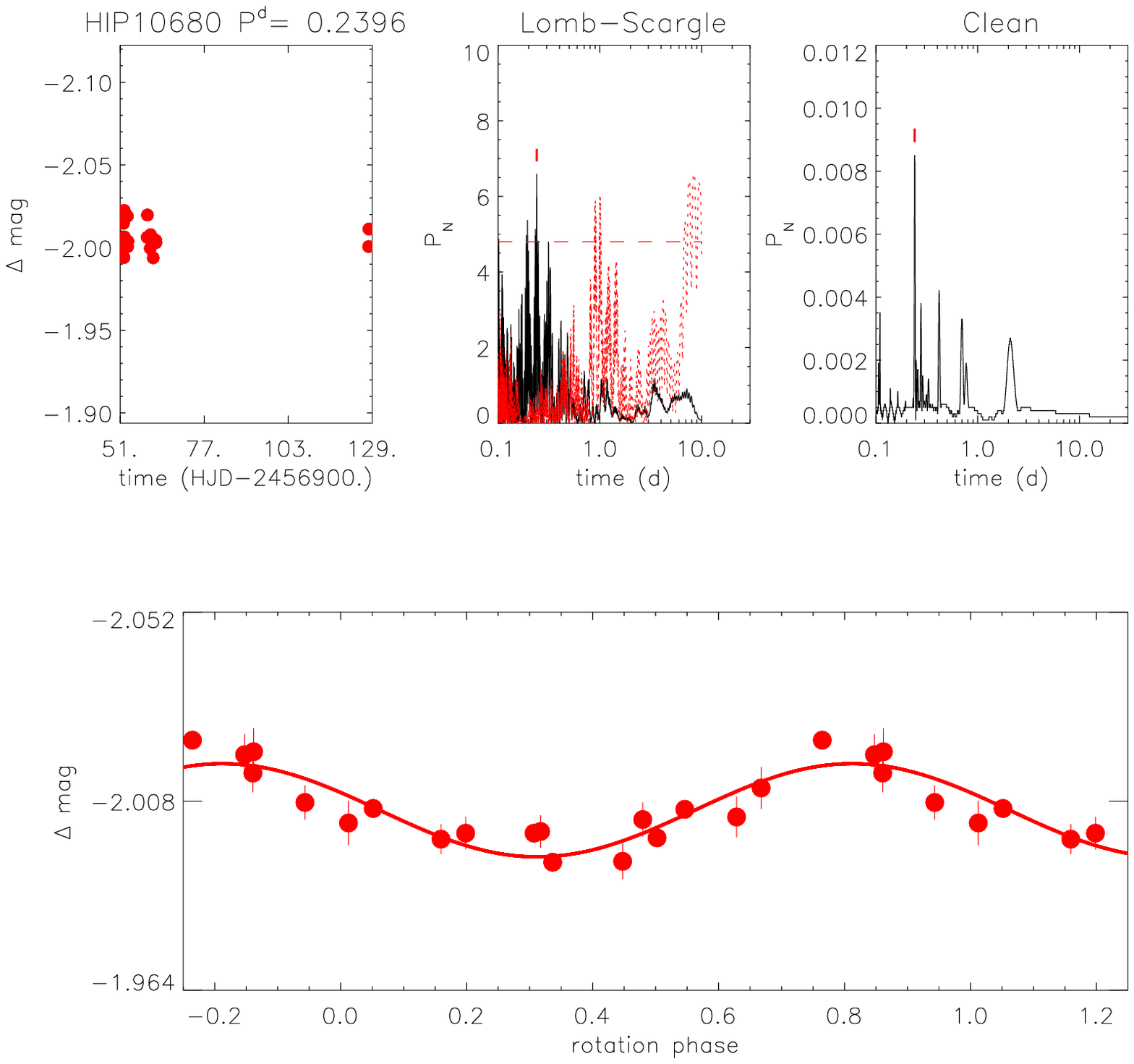}{\it top panels: \rm (left) Our new observations (combined B, V, and R magnitudes; see text) of 
HIP\,10680 collected at the Taurus Hill Observatory; (middle) LS periodogram (dotted line is the window function and 
horizontal dashed line the power corresponding to a  99\% confidence level); (right) CLEAN periodogram. 
\it bottom panel: \rm light curve phased with the  P = 0.2396d rotation period and with overplotted (solid line) a sinusoidal 
fit with an amplitude of $\Delta$mag = 0.026 mag.}
\IBVSfigKey{messina_fig1.eps}{HIP10680}{hip10680}

 HIP\,10680 is reported in the Hipparcos catalogue  as  likely algol-type eclipsing binary with  period P = 7.06 d. However, a note to the catalog reports the possibility that this photometry has been contaminated at some epochs by the presence of the close companion generating a spurious variability. 

Consistently with the young age and their low-mass, we expect that both components exhibit photometric variability,
possibly periodic, caused by the presence of surface temperature inhomogeneities. The photometric variability can in principle
allow us to measure the rotation period. Multi-band photometric observations are suited to infer the rotation period and 
can add information on the nature of surface
inhomogeneities, i.e. on their temperature, and on a lower limit on their covering fraction.

\section*{Observations}
To measure the photometric rotation periods of both components we carried out a multi-filter photometric 
monitoring at the Taurus Hill Observatory (62$^{\circ}$\,18$^{\prime}$\,54$^{\prime\prime}$N and 
28$^{\circ}$\,23$^{\prime}$\,21$^{\prime\prime}$E, 160\,m a.s.l, Varkaus, Finland). Observations were collected 
with a  35-cm f/11 Celestron telescope on a Paramount ME German equatorial mount, and equipped with a 
SBIG ST-8XME CCD camera (1530$\times$1020, 9\,$\mu$m pixels size), and  
Johnson-Bessell BVR filters.

The visual binary HIP\,10680/HIP\,10679 was observed from October 21, 2014 until January 5, 2015 for a total of 7 nights. 
We observed in the B, V, and R filters and collected a total of 90 frames in each filter. On a few nights, we observed the binary
up to four times at distance of about  2 hours from one pointing to the subsequent one. On each pointing, we collected five
consecutive frames per filter. Exposure times were set to 15, 6, and 2 sec for the B, V, and R filters,
respectively. Bias subtraction and flat fielding of science frames were performed with MaxIm DL 5.0 
(Diffraction Limited, Canada) and the magnitude timeseries of each binary's component and other nearby stars were extracted using 
aperture photometry. Each series of five consecutive magnitudes was averaged for the subsequent analysis. After averaging, 
we were left with 17 averaged magnitudes per filter whose photometric precisions turned out to be $\sigma_{\rm B}$ = 0.006, 
$\sigma_{\rm V}$ = 0.006, and $\sigma_{\rm R}$ = 0.007 mag.
The stars BD+28 381 (RA = 02:17:10.77, DEC =  +28:40:55.60, J2000.0, V = 9.09, B$-$V = 1.06) and GSC\,1777-01383 (RA = 02:17:24, DEC = 28:40:39, J2000, V = 12.82)
turned out to be well suited to be used as comparison (C) and check (CK) stars to get differential magnitudes of our targets. The standard deviation of the CK$-$C
magnitude time series turned out to be $\sigma_{CK-C}$ =  0.009\,mag.\\
\IBVSfig{15cm}{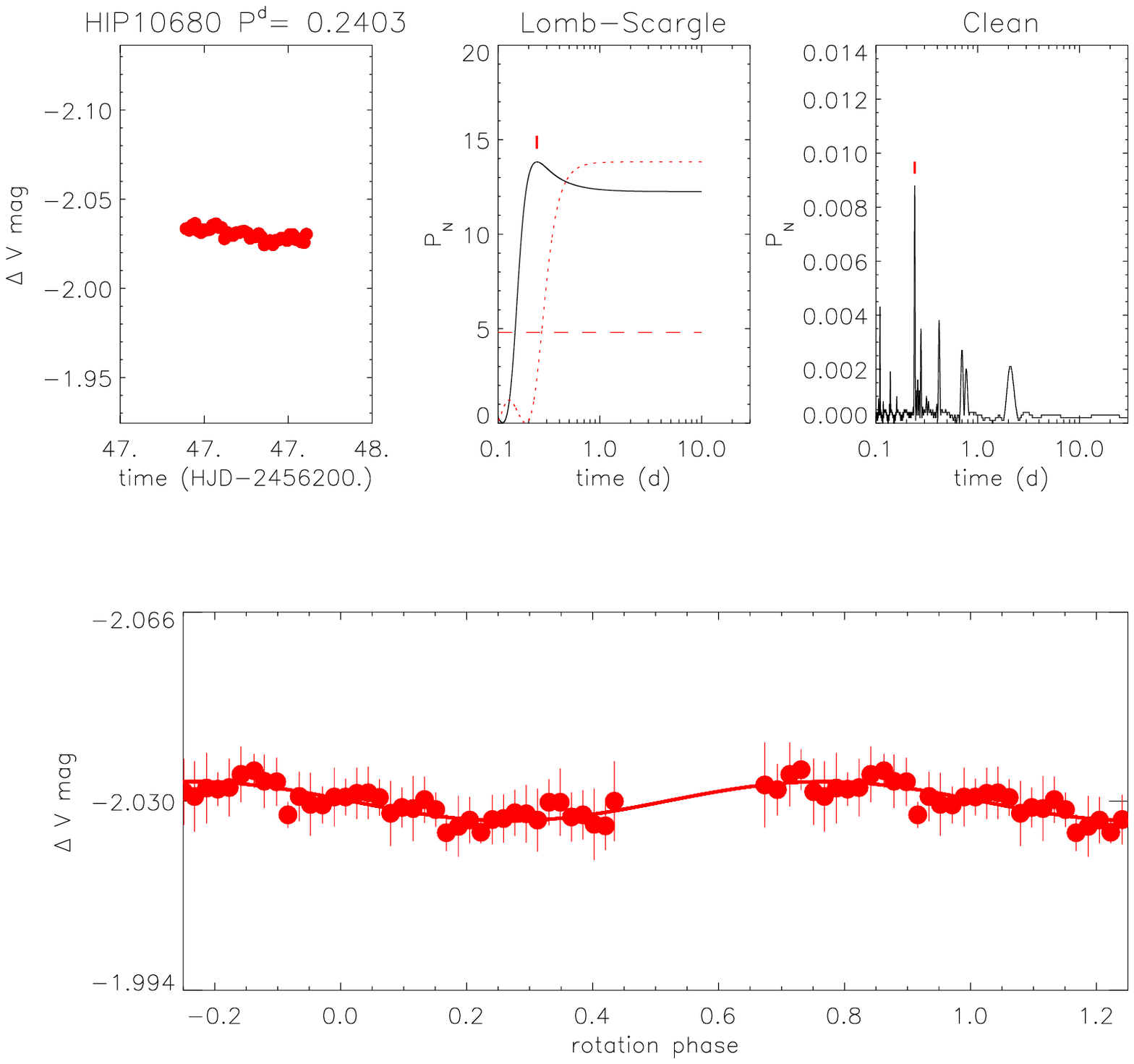}{The same as in Fig.\,1, but for data collected at Canis Mayor Observatory in the V band
and phased with the rotation period P = 0.2403\,d.}
\IBVSfigKey{messina_fig2.eps}{HIP10680}{hip10680_rob}
On one night, November 15, 2012, we could get a series of 390 frames in the V filter at the Canis Mayor Observatory
(44$^{\circ}$\,06$^{\prime}$\,17$^{\prime\prime}$N and 10$^{\circ}$\,00$^{\prime}$\,29$^{\prime\prime}$E, 190 m a.s.l., La Spezia, Italy).
Observations were collected by a 40-cm f/8 telescope equipped with a SBIG STL 6303 CCD camera (0.58$^{\prime\prime}$/pixel
plate scale and   29.5$^{\prime}$$\times$19.7$^{\prime}$ field of view) using 10-s exposure. Frame reduction was  done \rm as
already described for the data collected at  the Taurus Hill Observatory.

 \section*{Rotation period search}

\subsection*{HIP\,10680}
We carried out a Pearson linear correlation analysis among the magnitude variations in
different filters and found that B, V, and R magnitude variations were well correlated
(we measured the following linear correlation coefficients: r$_{\rm BV}$ =  0.61; r$_{\rm BR}$ =  0.54; r$_{\rm VR}$ =  0.57 with significance level
$>$ 99.9\%).
To improve the S/N ratio of the magnitude timeseries for the periodogram analysis, we averaged the B, V, and R band light curves. 
The Lomb-Scargle (LS; Scargle 1982) and CLEAN (Roberts et al. 1987) periodogram analyses revealed a significant (FAP $<$ 1\%) power peak at P = 0.2396$\pm$0.0005\,d
which we consider the stellar rotation period. For instance, this is to date the shortest rotation period ever measured in a member of the $\beta$ Pictoris association. 
The light curve amplitudes inferred from the amplitude of the
sinusoidal fit are $\Delta$B = 0.035, $\Delta$V = 0.026, $\Delta$R = 0.021 mag. 
An estimate of the False Alarm Probability
(FAP), that is the probability that a peak of given power in
the periodogram is caused by statistical variations, i.e., by
Gaussian noise, was done using Monte Carlo simulations ac-
cording to the approach outlined by Herbst et al. (2002).
The uncertainty on the rotation period determination was
estimated following Lamm et al. (2004; see also Messina et
al. 2010).
The results are summarized in Fig.\,1.

The results of the periodogram analysis of the data collected at the Canis Mayor Observatory are summarized in Fig.\,2.
In this case, we note that the observations lasted about 0.19\,d, and, therefore, were not long enough
to measure the rotation period of HIP\,10680 (the time span of observations should be at least longer than 1.5 times the searched rotation period).
Nonetheless, thanks to the very high sampling we could retrieve the correct rotation period and, consistently with the other datasets, we
presented  the same analysis. In this case the results can be considered as a confirmation
rather than an independent determination of the rotation period of HIP\,10680. 

We could retrieve observations of this binary system also from the SuperWASP (Butters et al.
2010) and Hipparcos  (Turon et al. 1993) public archives.\\
This binary system was observed by SWASP (1SWASP\,J021725.28 +284442.1) on three nights only, from 19 to 21 July, 2008.
A total of 21 V-band frames were collected, where the two components are not spatially resolved. Owing to the star's brightness,
the photometric precision was very high ($\sigma_V$ = 0.003 mag). The LS and Clean periodogram analyses revealed the most significant power peak 
at P = 0.2405\,d, which is in very good agreement with our independent period determination. Although the components are unresolved
in the SuperWASP photometry, the flux variability is likely dominated by the brighter F5V component (HIP\,10680). The results are summarized in Fig.\,3.

\IBVSfig{15cm}{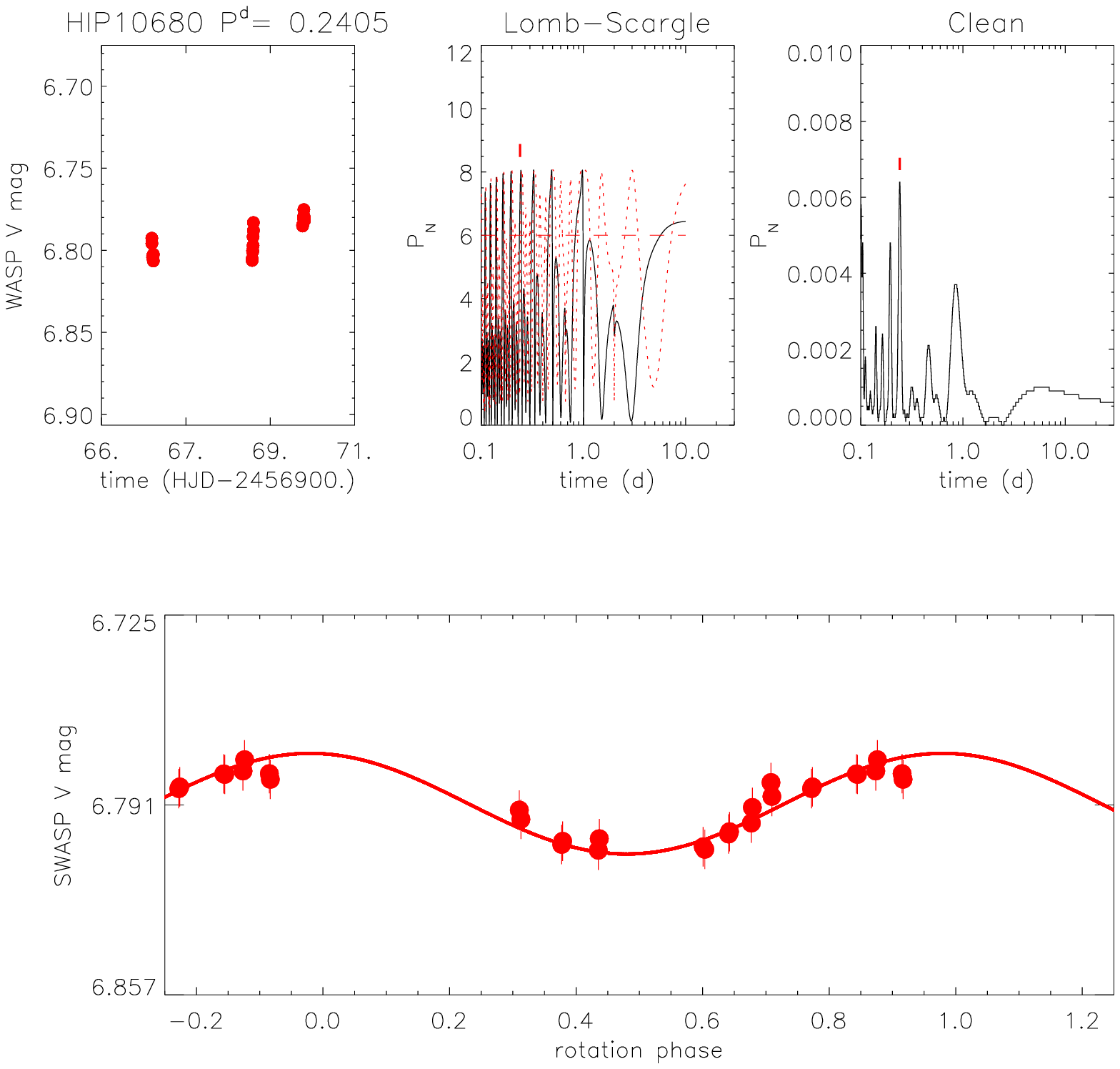}{The same as in Fig.\,1, but for data collected by SuperWASP for the unresolved system HIP10680+HIP10679. The 
light curve is phased with the  P = 0.240\,d rotation period and with overplotted 
(solid line) a sinusoidal fit with an amplitude of $\Delta$V = 0.035\,mag.}
\IBVSfigKey{messina_fig3.eps}{HIP10680}{hip10680_sw}

This binary system was observed also by Hipparcos  from January 1990 to March 1992.
After removing outliers, and averaging consecutive observations collected within 20 min, a
total of 33 magnitudes were left for the subsequent analysis. Owing to the star's brightness,
the photometric precision was very high ($\sigma_V$ = 0.007 mag). The LS and CLEAN periodogram analyses revealed the most significant power peak 
at P = 0.2805\,d, and P = 0.2005\,d. A note to the Hipparcos catalogue reports the possibility
that this photometry has been contaminated at some epochs by the presence of the close companion generating
a spurious variability. This may explain the about 10\% discrepancy with respect to the period derived from our own and the SuperWASP photometry.
The results are summarized in Fig.\,4.

\IBVSfig{15cm}{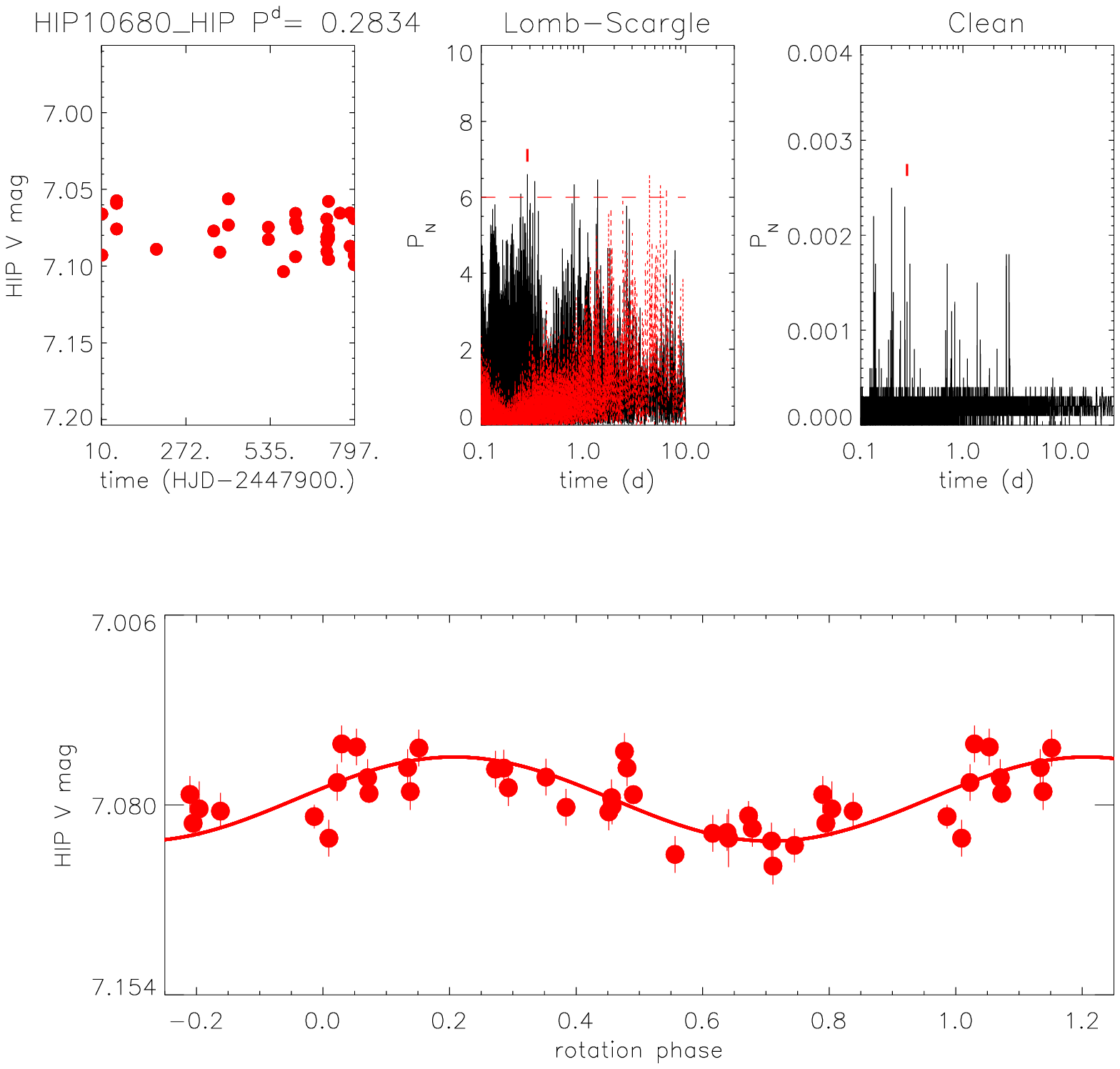}{The same as in Fig.\,1, but for data collected by Hipparcos. The 
light curve is phased with the  P = 0.240\,d rotation period and with overplotted 
(solid line) a sinusoidal fit with an amplitude of $\Delta$V = 0.030\,mag.}
\IBVSfigKey{messina_fig4.eps}{HIP10680}{hip10680_hip}

\subsection*{HIP10679}
We carried out a Pearson linear correlation analysis among the magnitude variations in
different filters and found that the correlation coefficients are $r$ $>$ 0.70 with confidence level $>$ 99.8\%.
As done for HIP\,10680,  we averaged the multi-band light curves. The LS and CLEAN
periodograms revealed the highest power peak to be at P = 0.777$\pm$0.005d 
with FAP $<$ 1\%. This is the stellar rotation period 
of HIP\,10679. The light curves have peak-to-peak amplitudes $\Delta$B = 0.06, $\Delta$V =0.07, and $\Delta$R = 0.07 mag.
The results are summarized in Fig.\,5.

We could retrieve also  the magnitude timeseries of HIP\,10679 collected by Hipparcos. Although, the magnitudes are to some level
contaminated by
the flux from the nearby brighter star, we could retrieve from our periodogram analysis  about the same rotation period P =  0.78$\pm$0.02\,d.
The results are summarized in Fig.\,6.
No similar rotation period was found in the short SuperWASP timeseries. \\

 \section*{Discussion}

Using the observed V magnitude, the distance from Pecaut \& Mamajek (2013), the bolometric correction and effective temperature proper for their spectral types from Pecaut \& Mamajek (2013)
we could estimate the luminosity and radius of both components. 
For HIP\,10680, we derive a luminosity L = 1.88$\pm$0.17 L$_\odot$, a radius R = 1.11$\pm$0.10 R$_\odot$. Combining radius and average projected stellar velocity,
we estimate an inclination of the stellar rotation axis $i$ $\sim$ 10$^{\circ}$.\\
For HIP\,10679, we derive a luminosity L = 0.96$\pm$0.09 L$_\odot$, a radius R = 0.95$\pm$0.09 R$_\odot$. Combining radius and average projected stellar velocity we estimate an inclination of the stellar rotation axis $i$ $\sim$ 10$^{\circ}$ .
The same inclination likely arises from the common formation and early evolution processes of the two stars in the same binary system.
An interesting aspect presented by this system is that the two components have a significant difference in their rotation periods. 
This difference may be due to the different masses. However, we find from a comparison with the evolutionary models of Siess et al. (2000) that this difference
is not larger than about 15\%. Different initial rotation periods may also have caused the
presently observed difference. However, we note that the slower rotating G2V component hosts a debris disc. 
There is evidence of an anti-correlation between the presence of IR excess, revealing the presence of primordial discs, 
and the rotation period in very young stars (see, e.g. Bouvier et al. 1993, Rebull et al. 2004). In fact, the magnetic disc-locking
should lock the rotation of the external star's envelope with the disc rotation and prevent the star to spin-up despite the stellar radius contraction.
By the age of $\beta$ Pictoris, such an anti-correlation is not as significant as in younger stars, and it appears as a weak tendency of fast rotators
to have smaller IR excess (see Rebull et al. 2008). However, the available sample is not large and $v\sin{i}$ is used to measure the rotation rate,
instead of the more robust rotation period.
In our specific case, one possibility to explain the rotation period difference is that the component with IR excess HIP10679
may have had a disc-locking phase longer than the other component, for which no IR excess is detected.
The shorter disc-locking phase of HIP10680 may have allowed this star to start the rotation spin-up, owing to radius contraction towards the ZAMS,
earlier than HIP10679, and therefore reaching a shorter rotation period at the present age. However, we just propose it as one possibility.\\  
What may have caused different disc-lifetimes for the two components and different rotation periods is currently unknown. \rm In fact, neither binarity nor the presence
of sub-stellar companion have been reported for both stars, that may have gravitationally perturbed the primordial disc of HIP\,10680, enhancing its dispersal.

\IBVSfig{15cm}{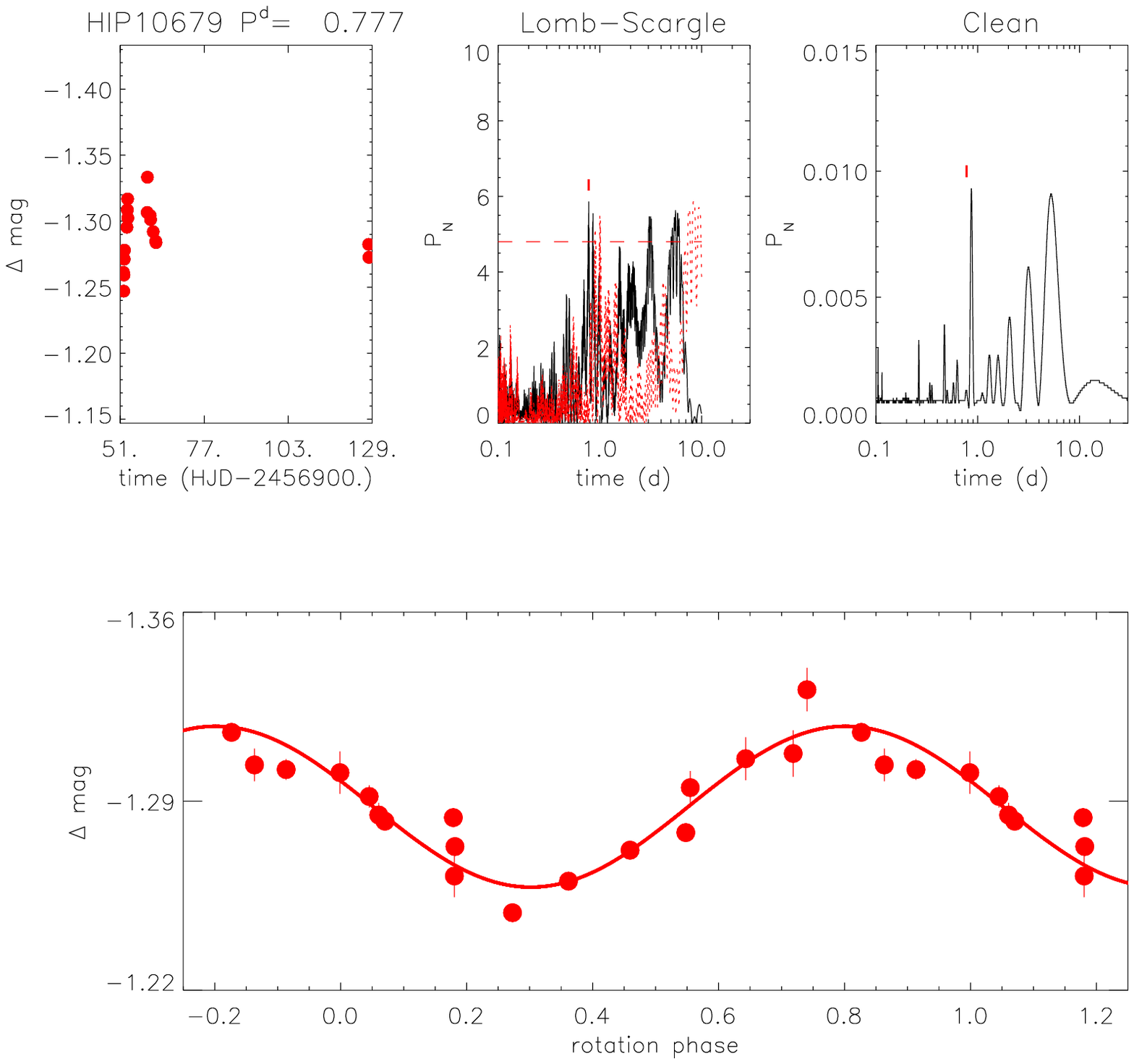}{\it top panels: \rm Our new  observations (combined B, V, and R magnitudes; see text) of HIP\,10679; LS periodogram (dotted line is the window function and horizontal dashed line the power corresponding to a 
99\% confidence level); and CLEAN periodograms. 
\it bottom panel: \rm light curve phased with the  P = 0.777\,d rotation period and with overplotted 
(solid line) a sinusoidal fit with an amplitude of $\Delta$mag = 0.07\,mag.}
\IBVSfigKey{messina_fig5.eps}{HIP10679}{hip10679}

\IBVSfig{15cm}{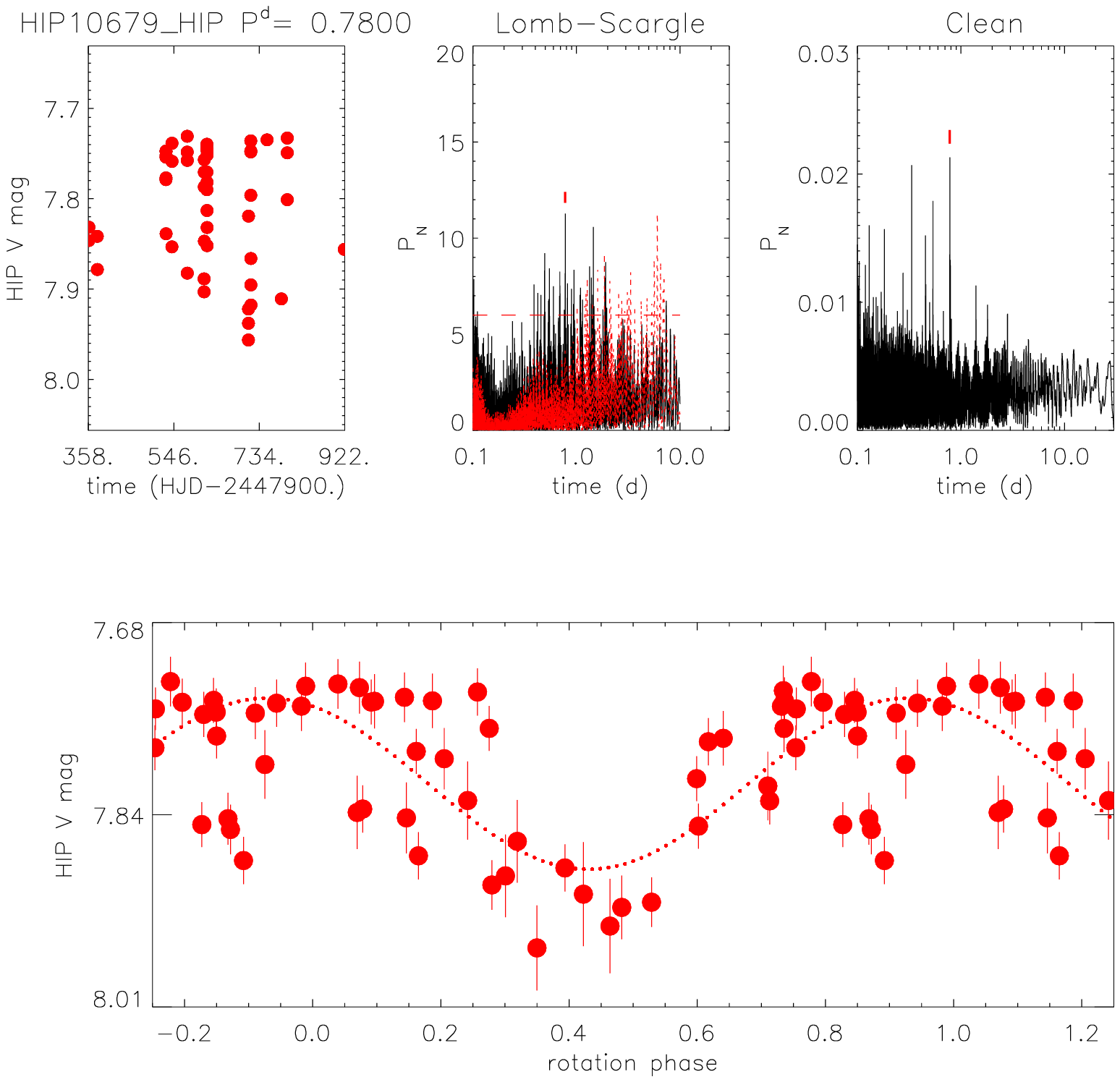}{The same as in Fig.\,5, but for data collected by Hipparcos. A mentioned in the text,
this photometry may be contaminated by the flux from the brighter component.}
\IBVSfigKey{messina_fig6.eps}{HIP10679}{hip10679_hip}

\section*{Conclusions}
We have carried out a multi-filter photometric monitoring of the wide visual binary HIP10680/HIP10679.
We found that HIP10680 has a rotation period  P = 0.2396$\pm$0.0005\,d, which is the shortest ever measured in the $\beta$ Pictoris association,
whereas HIP10679  has a rotation period  P = 0.777$\pm$0.005\,d. Combining stellar radii and projected rotational velocities,
we found that both components have same inclinations of their rotation axes, $i$ $\sim$ 10$^{\circ}$ and,
therefore, they are seen almost pole-on. Despite the low inclination, both components exhibit a significant
photometric variability whose amplitudes in the V band are $\Delta$V = 0.03 mag and $\Delta$V = 0.07 mag,
for HIP10680 and HIP10679, respectively. The G2V star, having a deeper convection zone, and consequently,
a more efficient dynamo action, shows a larger amplitude variability.
Although the two components have a mass difference not larger than 15\%, they exhibit a 
significant difference between their rotation periods. Such difference may arise either from different initial rotation periods or to different disc life times.
For instance, the slower component HIP\,10679 hosts a well know debris disc.\\

{\bf Acknowledgements:}
The extensive use  of the SIMBAD  and ADS  databases  operated by  the  CDS center,  Strasbourg,
France,  is gratefully  acknowledged. We thank the Super-WASP consortium for the use of their public archive in this research. 
We also thanks the anonymous Referee for useful comments and suggestions.

\references
Bouvier, J., Cabrit, S., Fernandez, M., Martin, E. L., \& Matthews,
J. M. 1993, A\&A, 272, 176

Brandt, T.D., Kuzuhara, M., McElwain, M.W. et al. 2014, ApJ, 786, 1

Butters, O.W., West, R.G., Anderson, D.R., et al., 2010, A\&A, 520,
L10
	
da Silva, L. Torres, C.A.O., de la Rez, R., et al. 2009, A\&A, 508, 833

Herbst, W., Bailer-Jones, C. A. L., Mundt, R., Meisenheimer, K., \&
Wackermann, R. 2002, A\&A, 396, 513
	
%Hog, E., Fabricius, C., Makarov, V.V., et al. 2000, A\&A, 355, 27

Kiss, L. L., Moor, A., Szalai, T. et al. 2011, MNRAS, 411,878

Lamm, M. H., Bailer-Jones, C. A. L., Mundt, R., Herbst, W., \&
Scholz, A. 2004, A\&A, 417, 557

Lepine, S. \& Simon, M., 2009, AJ, 137, 3632

Malo, L., Doyon, R., Lafreniere, D. et al. 2014, ApJ, 762, 88

Mamajek, E.E. \& Bell, Cameron P. M. 2014, MNRAS, 445, 2169

Mason, B.D., Wycoff, G.L., Hartkopf, W.I., et al. 2001, AJ,122, 3466

Mentuch, E., Brandeker, A., van Kerkwijk, M.H. et al. 2008, ApJ, 689, 1127

Messina, S., Desidera, S., Turatto, M., Lanzafame, A. C., \& Guinan,
E. F. 2010, A\&A, 520, A15

Messina, S., Monard, B., Biazzo, K., Melo, C. H. F., \& Frasca, A.
 2014, A\&A, 570, A19
 
Messina, S., Monard, B., Worters, H.L.,  Bromage, G.E.,  Zanmar, R.S. 2015, in press by New Astronomy

Rebull, L. M., Wollfs S. C., \& Strom, S. E. 2004, AJ, 127, 1029

Rebull, L. M., Stapelfeldt, K. R., Werner, M. W. et al.  2008, ApJ, 681, 1484

Roberts, D. H., Lehar, J., \& Dreher, J. W. 1987, AJ, 93, 968

Pecaut, M. J. \& Mamajek, E. E. 2013, ApJS, 208, 9

Riviere-Marichalar, P., Barrado, D., Montesinos, B.,  et al. 2014, A\&A 565, A68

Scargle, J. D. 1982, ApJ, 263, 835

Siess L., Dufour E., Forestini M. 2000, A\&A, 358

Torres, C. A. O., Quast, G. R., da Silva, L., et al. 2006, A\&A, 460, 695

Turon C., Egret D., Gomez A., et al. 1993, Bull. Inf. Centre Donnees Stellaires, 43, 5 

Valenti, J.A. \& Fischer, D.A. 2005, ApJS, 159, 141

Zacharias N., Finch C.T., Girard T.M., et al. 2013,  Astron. J., 145, 44 

Zuckerman, B. \& Song, I. 2004, ARA\&A, 42, 685

\endreferences

\IBVSedata{5xxx-t2.txt}
\IBVSedata{5xxx-t3.txt}

\IBVSefigure{5xxx-f2.ps}
\IBVSefigure{5xxx-f3.ps}
\IBVSefigure{5xxx-f4.ps}
\IBVSefigure{5xxx-f5.ps}

\end{document}